\documentclass[fleqn,usenatbib,letters]{mnras}
\usepackage[T1]{fontenc}
\usepackage{ae,aecompl}
\usepackage{graphicx}   % Including figure files
\usepackage{amsmath}    % Advanced maths commands
\usepackage{amssymb}    % Extra maths symbols
\usepackage{float}
\newcommand{\PIMA}{$\cal P\hspace{-0.067em}I\hspace{-0.067em}M\hspace{-0.067em}A\hspace{-0.1em}$ }
\newcommand{\Cov}{\rm Cov}
\newcommand{\ntab}[2]{ \multicolumn{1}{#1}{#2} }
\newcommand{\Gaia}{{\it Gaia}}
\usepackage{xcolor}
\definecolor{Dred}{rgb}{0.312,0.070,0.070}
\definecolor{Dblue}{rgb}{0.070,0.070,0.312}

\newcounter{note}
\let\oldmarginpar\marginpar
\renewcommand\marginpar[1]{\-\oldmarginpar[\raggedleft\footnotesize #1]%
{\raggedright\footnotesize #1}}
\newcommand{\Note}[1]{\Rdb{#1}{\addtocounter{note}{1}%
\marginpar{\small\underline{\Rdb{Comm \arabic{note}}}}}}
\newcommand{\note}[1]{\Rdb{#1}}
\renewcommand{\note}{}\renewcommand{\Note}{} %% Uncomment it to disable cherry-color comments
\newcommand{\timezone}{-0500}
\newcommand{\Number}[1]{\ifnum#1<10\relax0\number#1\else\number#1\fi}
\newcommand{\isodate}{
\count151=\time
\divide\count151 by 60
\count151=\count151
\multiply\count151 by 60
\count152=\time
\advance\count152 by -\count151
\divide\count151 by 60
\count152=\count151
\multiply\count151 by 60
\count153=\time
\advance\count153 by -\count151
\Number{\year}.\Number{\month}.\Number{\day}--\Number{\count152}:\Number{\count153} \enskip \timezone
}

\title[On significance of VLBI/Gaia position offsets]{On significance of VLBI/Gaia position offsets}
\author[Petrov and Kovalev]{
L.~Petrov$^1$\thanks{E-mail: Leonid.Petrov@lpetrov.net (LP)},
and
Y.~Y.~Kovalev$^{2,3}$\\
$^1$Astrogeo Center, 7312 Sportsman Dr., Falls Church, VA 22043, USA\\
$^2$Astro Space Center of Lebedev Physical Institute, Profsoyuznaya 86/32, 117997 Moscow, Russia\\
$^3$Max-Planck-Institut f\"ur Radioastronomie, Auf dem H\"ugel 69, 53121 Bonn, Germany
}

\date{Accepted 2017 January 2; Received 2016 December 15; in original form 2016 November 6}

\pubyear{2017}

\begin{document}
\label{firstpage}
\pagerange{\pageref{firstpage}--\pageref{lastpage}}
\maketitle

\begin{abstract}
   We have cross matched the \Gaia\ Data Release 1 secondary dataset that
contains positions of 1.14 billion objects against the most complete
to date catalogue of VLBI positions of 11.4 thousand sources, almost
exclusively active galactic nuclei. We found 6,064 matches, i.e. 53\% radio
objects. The median uncertainty of VLBI positions is a factor of 4 smaller
than the median uncertainties of their optical counterparts. Our analysis 
shows that the distribution of normalized arc lengths significantly 
deviates from Rayleigh shape with an excess of objects with small 
normalized arc lengths and with a number of outliers. We found 
that 6\% matches have radio optical offsets significant at 99\% confidence 
level. Therefore, we conclude there exists a population of objects with 
genuine offsets between centroids of radio and optical emission.
\end{abstract}


\section{Introduction}

%% \par\vspace{-96ex} \fbox{\LARGE\it Draft of \isodate. \enskip \bf CONFIDENTIAL} \par \vspace{91ex}\par 

  The secondary dataset of the first release of astrometric data from
the European Space Agency mission \Gaia\ contains positions of 1.14 billion 
objects \citep{r:gaia_dr1}. Of them, the vast majority are stars, though 
over one hundred thousands of extragalactic objects, namely active galactic 
nuclei (AGN), were also included in the catalogue. The position uncertainty
of the \Gaia\ DR1 secondary dataset, 2.3~mas median, is two orders of magnitude 
higher than the uncertainty
of previous large all-sky catalogue in optical wavelengths NOMAD 
\citep{r:nomad} of 1.17 billion objects. The only technique that can 
determine positions of target sources with comparable accuracy is very 
long baseline interferometry (VLBI). The first insight on 
% \citep[e.g.,][]{r:takahashi00
comparison of \Gaia\ and VLBI position catalogues can be found in 
\citet{r:gaia_icrf2}, who found that the overall agreement between the 
optical and radio positions is excellent, though a small number of sources
($\sim\! 6$\%) show significant offsets. 
%This paper gives a good start for 
%further investigation of radio and optical offsets.

  In this letter we make our own comparison of \Gaia\ and VLBI positions
beyond that reported in \citet{r:gaia_icrf2}. Several factors motivated us. 
Firstly, the authors of the cited paper ran their comparison against the 
auxiliary \Gaia\ quasar solution for some 135,000 quasars. This 
solution is not yet published in full, and only positions of 2\% of the
objects were reported. The question of how results of the comparison against 
this auxiliary solution are representative to the main solution 
of one billion objects remained opened. 

  Secondly, \citet{r:gaia_icrf2} used the ICRF2 catalogue \citep{r:icrf2}
for their comparison. This catalogue assembled in 2008--2009 represented 
the state of the art by 2008 and comprised of sources observed in geodetic 
programs \citep{r:icrf1,r:geo_vlba} and six Very Long Baseline Array (VLBA) 
Calibrator Surveys \citep{r:vcs1,r:vcs2,r:vcs3,r:vcs4,r:vcs5,r:vcs6}. Since
that, there was an explosive growth of absolute astrometry VLBI programs:
VLBA and European VLBI network Galactic plane surveys \citep{r:vgaps,r:egaps};
VLBA Imaging and Polarimetry Survey (VIPS) \citep{r:astro_vips}; Australian
Long Baseline Calibrator Survey (LCS) \citep{r:lcs1}; the VLBA Calibrator 
Search for the BeSSel Survey \citep{r:bessel_search}; the VLBA survey of bright 
2MASS galaxies (V2M) \citep{r:v2m}; the VLBA+EVN survey of optically bright 
extragalactic radio sources (OBRS--1,OBRS--2) \citep{r:obrs1,r:obrs2}; the 
second epoch VLBA calibrator survey observations (VCS-ii), \citep{r:vcs-ii}. 
Besides, there is a number of ongoing surveys: the VLBI Ecliptic Plane Survey 
(VEPS) \citep{r:veps}, the wide-band VCS7,VCS8,VCS9 surveys \citep{r:vcs9}, 
and the VLBI survey of {\it Fermi} detected $\gamma$-ray sources 
\citep{r:faps2}. By September 14, 2016, the date of \Gaia\ DR1 release, the 
total number of sources with positions determined with absolute astrometry 
using VLBI reached 11,444, a factor of 3.5 increase with respect to the ICRF2. 

\newpage

  Thirdly, the analysis of \citet{r:gaia_icrf2} showed that there exist sources 
with significant radio-optical offsets. Early comparisons of source positions 
from VLBI and ground optical observations prompted \citet{r:zacharias14} and 
\citet{r:orosz13} to surmise there is a population of radio optical offset 
objects with position differences in a range of 10--100~mas. Large offsets can 
occur either due to unaccounted errors in optical positions or a gross 
oversight in deriving VLBI coordinates, or due to an offset between the centroids 
of radio and optic emission. We call latter objects genuine radio optical offset 
(thereafter, GROO) sources. An increase in the accuracy of the optical positions 
by two orders magnitude allows us to re-examine the question of the GROO 
population existence. If such a population exists, it poses a challenge to 
explain significant radio-optical offsets.

\section{Association of VLBI and Gaia objects}

  Our study is based on the analysis of the catalogue called \Gaia\ DR1 
secondary dataset. We used for our work positions, their uncertainties, 
correlations between right ascension and declination for 1,142,679,769 
objects. We did not analyze \Gaia\ data and took the catalogue as it is. 
On the other hand, we reprocessed all publicly available VLBI data listed 
in the previous section from the level of visibilities using VLBI data 
analysis software \PIMA\footnote{See \href{http://astrogeo.org/pima}{http://astrogeo.org/pima}}. 
Detailed description of the analysis strategy and comparison with
the methods adopted in the past and those used for processing the data can 
be found in \citet{r:vgaps}. An important conclusion of that comparison was 
%
%, although processing with \PIMA\ indeed improves position accuracy, 
%especially for weak sources by means of elimination of known deficiencies 
%in the data analysis strategy used in the past, 
%
that it does not introduce 
systematic differences at least above the 0.2~mas level with respect to 
the old processing pipeline. The specific VLBI catalogue used in our study 
is rfc\_2016c\footnote{Available online at \href{http://astrogeo.org/rfc}
{http://astrogeo.org/rfc}}. It is based on all geodesy and absolute 
astrometry VLBI data since April 1980 through July 2016, including all
observations used for deriving the ICRF2 catalogue and those that became
publicly available since 2008.
%
%  
%
% Although there was a number of improvements in 
% modeling radio interferometry observations absent in the past, notably 
% the elaborated mass loading model \citep{r:malo} and computation of path 
% delay in the neutral atmosphere using the output of numerical weather 
% models \citep{r:pd}, their impact on results is small and can be ignored
% for the purposes of the present study. VLBI data analysis is traditionally 
% performed in the accumulative mode. That means a new solution is obtained
% using all prior observations, including all those used for deriving the ICRF2 
% catalogue. Thus, the main difference between VLBI catalogues rfc\_2016c 
% and the ICRF2 is that we reanalyzed all prior VLBA data and used new 
% observations that became public since 2008. 

  At the first step we identified all \Gaia\ sources that lie within $5''$ 
of VLBI objects and found 6954 preliminary matches. We should note the source 
density of \Gaia\ DR1 is substantially heterogeneous (see Figure~9 in 
\citealt{r:gaia_dr1}): the density in the Galactic plane exceeds by two order 
of magnitude the density near the Galactic poles. To take into account 
variations of \Gaia\ spatial source density, we counted \Gaia\ sources on 
the regular $0.25^\circ \times 0.25^\circ$ grid and normalized 
the count to the number of sources per steradian. Then for a given match we 
computed the probability of false association (PFA) as the product of local 
\Gaia\ source density and the area $\pi d^2$ where 
$d = L_{VG} + 3 \max(\sigma_{\rm g,maj},\sigma_{\rm v,maj}$), 
$L_{VG}$ is the arc length VLBI/\Gaia, $\sigma_{\rm g,maj}$ and 
$\sigma_{\rm v,maj}$ are semi-major error ellipse axes for \Gaia\ and VLBI 
respectively. This conservative estimate of the PFA takes into account 
possible errors that affect $d$ and represents rather its upper limit. 
The total number of matches with the PFA less than $2 \cdot 10^{-4}$ is 
6064. Of them, 9 are radio stars. We have excluded them from further analysis. 
According to the selected PFA cutoff criterion, the mathematical expectation 
of the number of spurious matches within our conservative sample is 0.03, 
i.e. less than one object. We certainly missed some real matches, but for 
the purpose of this letter it is more important to prevent false matches 
in the sample.

\begin{figure}
   \begin{center}
       \includegraphics[width=0.40\textwidth]{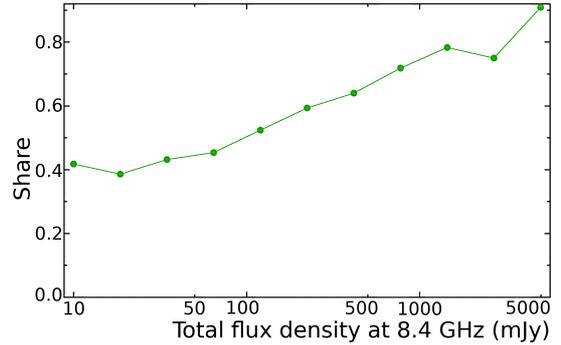}
   \end{center}
   \caption{The fraction of sources found in \Gaia\ catalogue as a function
            of the total flux density at 8.4 GHz integrated over 
            parsec-scale image in logarithmic scale.}
   \label{f:share_ass}
\end{figure}

  In total, 53\% VLBI sources are associated with a \Gaia\ counterpart.
The fraction of VLBI/\Gaia\ matches monotonically decreases with a decrease
of radio flux density: from 0.8 for sources with flux density $> 1$~Jy at 8~GHz
to 0.4 for sources with flux density in range of 10--40~mJy --- see 
Figure~\ref{f:share_ass}. At the same time, the diagram flux density versus 
G magnitude does not show any correlation. Since according to \cite{r:gaia_dr1}, 
the \Gaia\ DR1 is not complete in any sense, we defer analysis why 
the share of VLBI/\Gaia\ matches drops with a decrease of flux density 
till deep optical surveys, such as Pan-STARRS that is expected to be 
complete at least to 23~mag, will become available.

\begin{figure*}
    \includegraphics[width=0.40\textwidth]{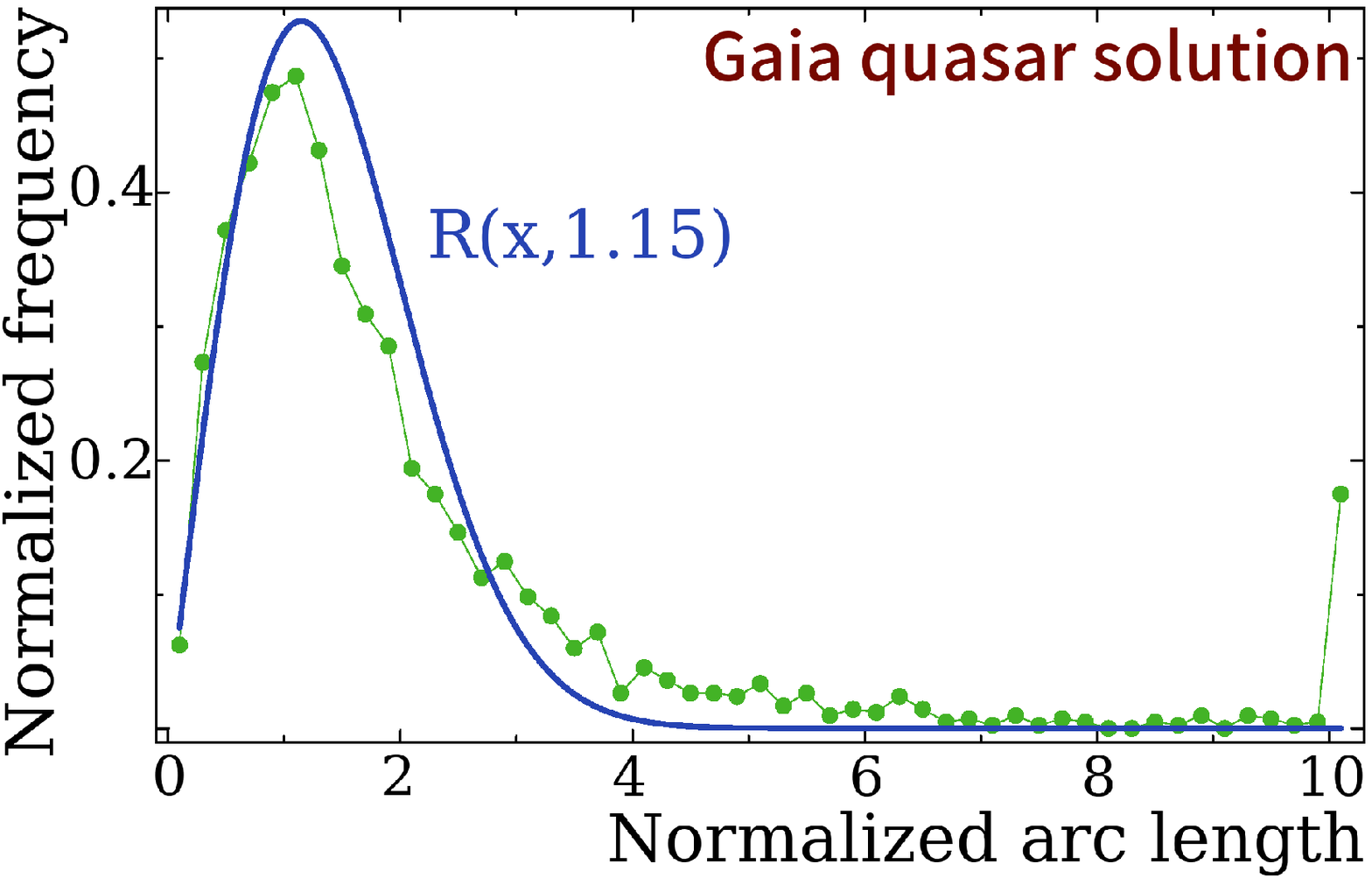}
    \includegraphics[width=0.40\textwidth]{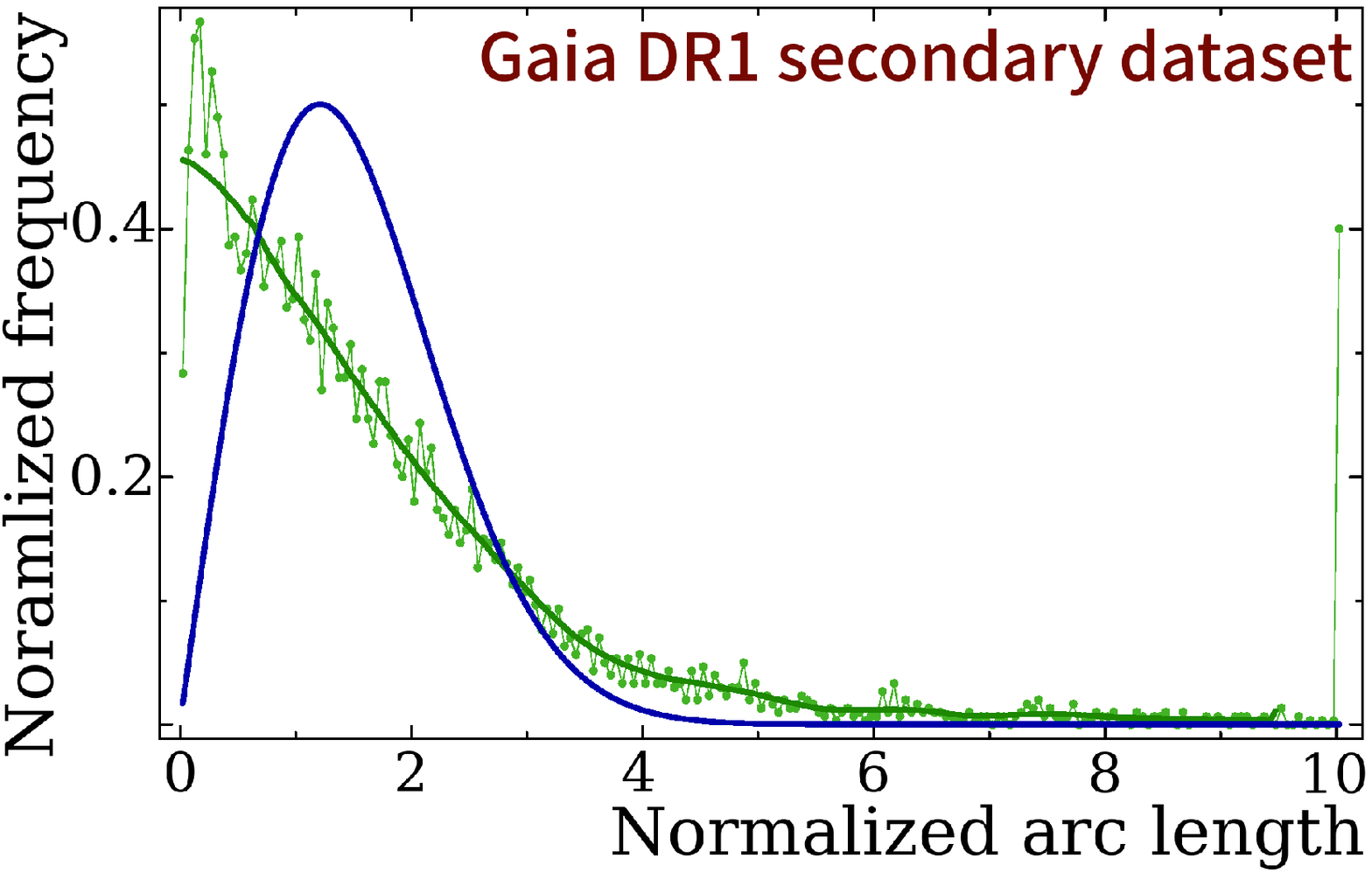}
    \caption{{\it Left:} normalized arc length among the 2080 VLBI/\Gaia\ 
             matches of QS sub-sample from the \Gaia\ quasar solution. 
             The continuous blue line shows the best fit Rayleigh
             distribution with parameter $\sigma= 1.15$. {\it Right:} 
             the distribution of normalized arc lengths among all 6055 
             matches from the \Gaia\ DR1 solution. The thick blue line 
             is the best fit to the Rayleigh distribution, which 
             is certainly inadequate.}
    \label{f:distr_orig}
\end{figure*}

\section{Analysis of VLBI/Gaia arc lengths}

   We computed the normalized arc lengths between VLBI positions from
rfc\_2016c solution and the \Gaia\ auxiliary quasar solution. We normalize 
the arc lengths exactly the same way as \citet{r:gaia_icrf2}:
$$
\begin{array}{l}
    q^2 = (d_\alpha, d_\delta)  \cdot \vspace{0.5ex} \\ 
          \left(
             \begin{array}{ll}
                \sigma^2_{g,\alpha} + \sigma^2_{v,\alpha}  & \Cov(\alpha,\delta)_g +
                                                             \Cov(\alpha,\delta)_v  \\
                \Cov(\alpha,\delta)_g + \Cov(\alpha,\delta)_v
                & \sigma^2_{g,\delta} + \sigma^2_{v,\delta}  
             \end{array}
          \right)^{-1} \cdot \vspace{0.5ex} \\
          (d_\alpha, d_\delta)^{^{\bf \top}} \,,
\end{array}
\label{e:e1}
$$
  where $d_\alpha, d_\delta$ are VLBI/\Gaia\ offsets in right ascension 
multiplied by factor $\cos{\delta}$ and declination,
$\sigma_{g,\alpha}$ and $\sigma_{v,\alpha}$ are reported uncertainty
in right ascensions (including the factor $\cos{\delta}$) of \Gaia\ and 
VLBI positions respectively, and $\sigma_{g,\alpha}$, $\sigma_{v,\alpha}$
are reported uncertainties in declinations.

  The distribution of normalized arc lengths, square root of $q^2$
of that sub-sample denoted as QS is shown in the left part of 
Figure~\ref{f:distr_orig}. The distribution is very close to that shown 
in Figure~8 of \citet{r:gaia_icrf2} based on analysis of the auxiliary 
\Gaia\ quasar solution and the ICRF2 catalogue. The blue line in the left 
part of Figure~\ref{f:distr_orig} shows the Rayleigh 
distribution\footnote{If position errors over each coordinate obey the 
Gaussian distribution, then the normalized arc lengths obey the Rayleigh
distribution.} with $\sigma = 1.15$ that fit best to the histogram that 
again is very close to the value 1.11 reported by \citet{r:gaia_icrf2}. 
This confirms our previous assertion that the differences in positions 
of sources common for the ICRF2 and rfc\_2016c catalogues are not 
essential for the present study. 

  However, the distribution of normalized arc lengths between positions 
from the \Gaia\ DR1 secondary solution and VLBI is remarkably different 
(right part of Figure~\ref{f:distr_orig}). It is definitely very far 
from the Rayleigh distribution.

\begin{figure}
    \centerline{\includegraphics[width=0.40\textwidth]{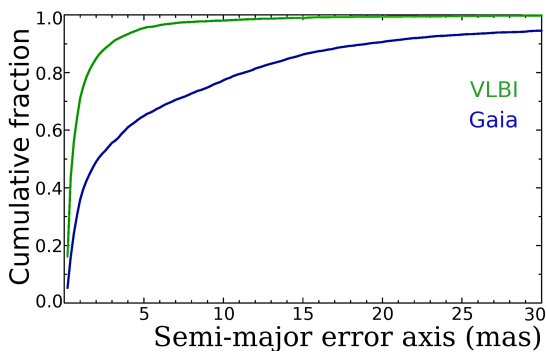}}
    \caption{Cumulative distribution function of semi-major error axes 
             $P(\sigma_{\rm maj}<a)$: green (upper) curve for VLBI and 
             blue (low) curve for \Gaia.}
    \label{f:cumul}
\end{figure}

   The normalized arc lengths depend on both arc lengths and uncertainties
in \Gaia\ and VLBI position estimates. Figure~\ref{f:cumul} demonstrate that
the \Gaia\ position errors dominate over VLBI positions in normalized arc 
lengths. In particular, the median $\sigma_{\rm v,maj}$ of the matches 
is 0.50~mas, while the median $\sigma_{\rm g,maj}$ is 2.15~mas (compare with 
2.3~mas for the total \Gaia\ DR1 sample), i.e. a factor of four greater. 
%Artificial scaling of VLBI uncertainties by a small number like 0.1 or by 
%a large number changes the distribution, but it still remains substantially 
%non-Rayleighian. 
The shape of the distribution remains non-Rayleighian even 
when we preform analysis of \Gaia\ DR1 and VLBI rfc\_2016c solutions among 
2088~matches of the QS sample. Therefore, we conclude the shape of the 
distribution is due to a peculiarity of the \Gaia\ DR1 secondary solution 
errors that did not affect strongly the \Gaia\ quasar auxiliary solution. 

   It is important to note that proper motions and parallaxes were estimated 
in the \Gaia\ secondary solution. Since the time span of the dataset used in 
producing the \Gaia\ DR1 solution, 14 months, in general is not sufficient for 
providing good estimates of parallax and proper motions, constraints were 
applied.  The reciprocal weights of constraints were adjusted to make 
realistic errors of positions and parallaxes of stars \citep{r:mich15} that 
do have proper motions and parallaxes. This disfavored treatment of AGNs 
that have negligible parallaxes and proper motion. Estimating proper motions 
and parallaxes in addition to positions of AGNs inflated their formal 
uncertainties. See \citet{r:mich15} for further details.

\begin{figure*}
    \includegraphics[width=0.40\textwidth]{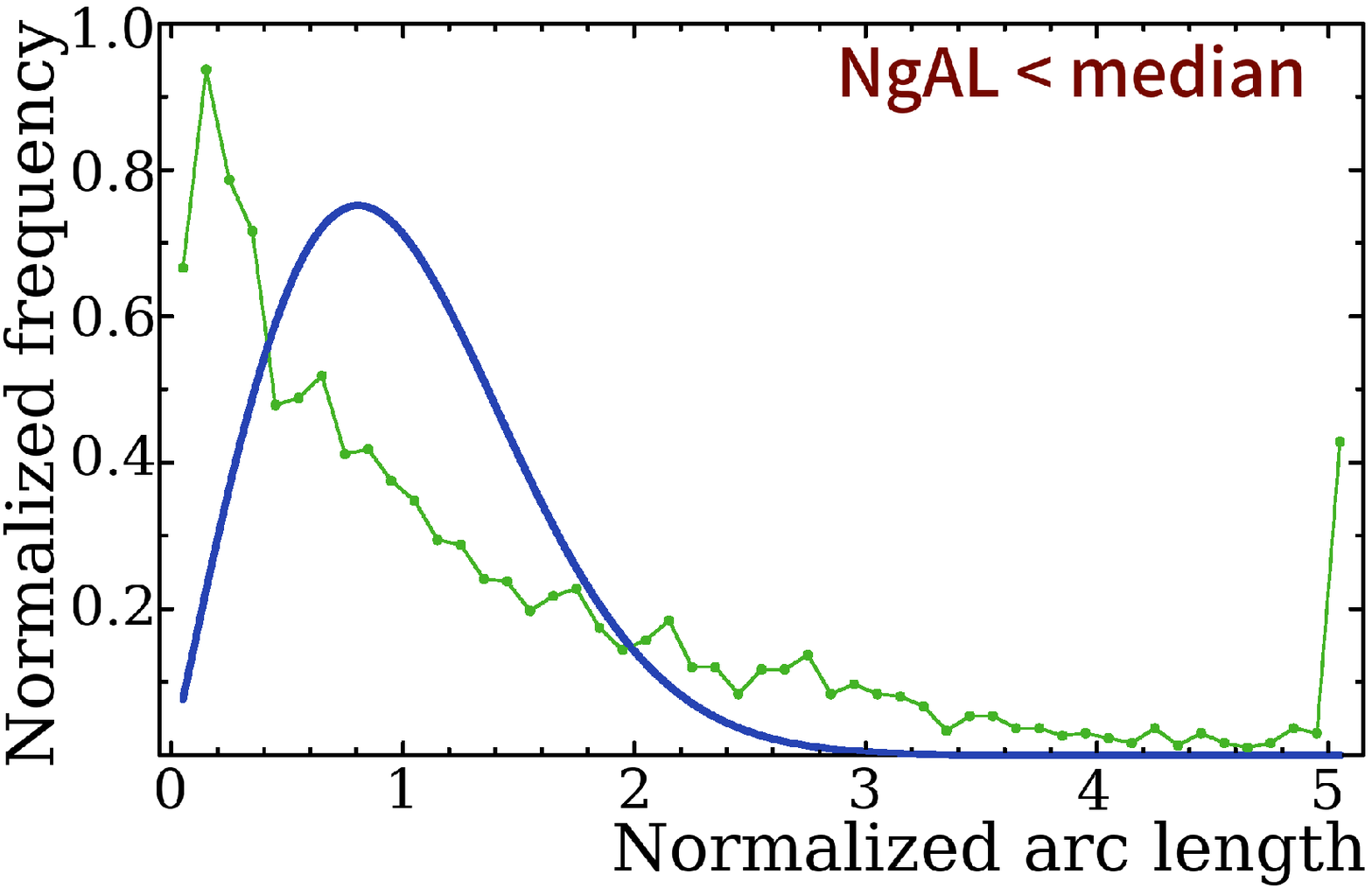}
    \includegraphics[width=0.40\textwidth]{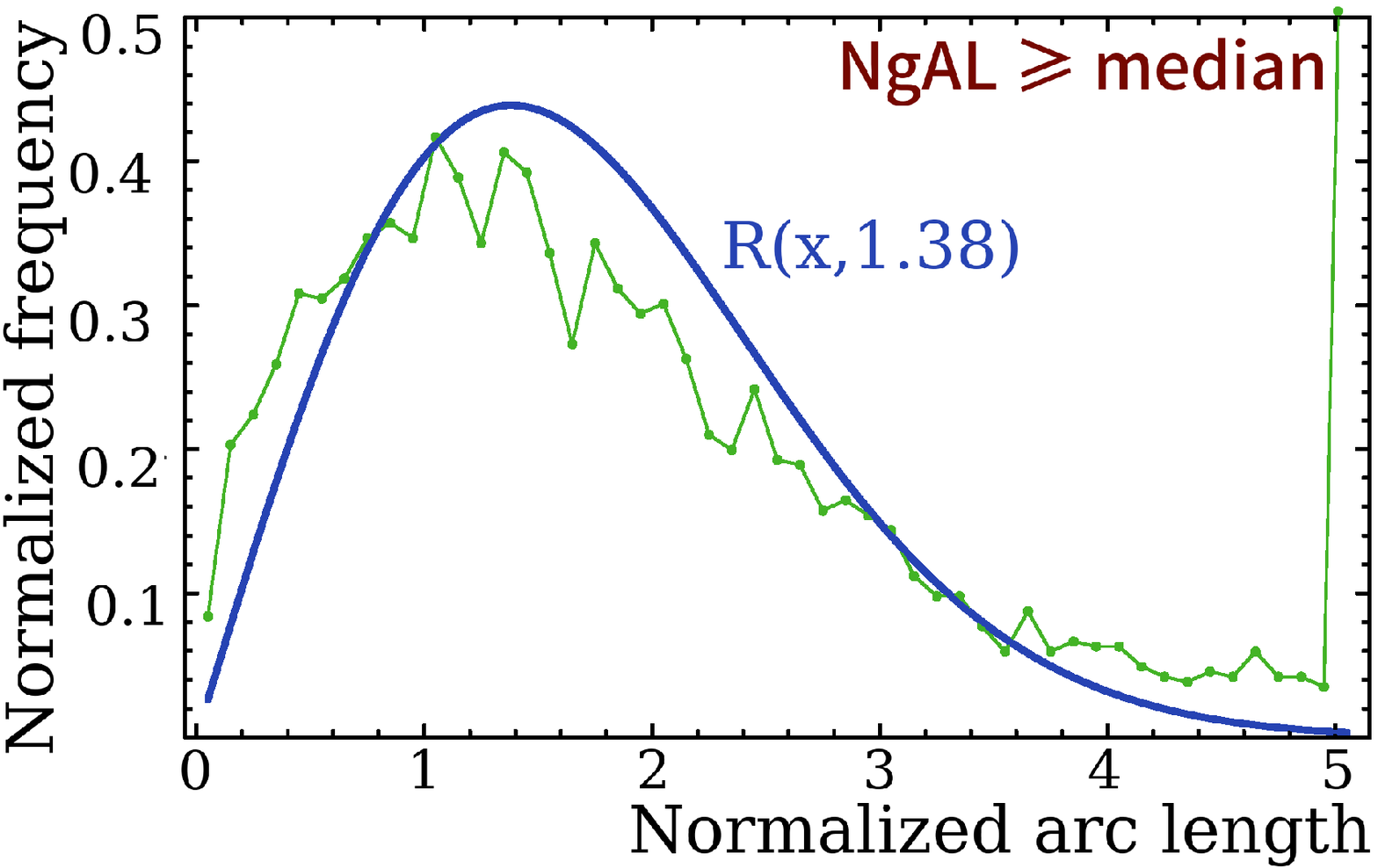}
    \caption{Normalized arc length distributions. {\it Left:} 
             the sub-sample of 3001 matches with the number of good 
             observations along scan below the median value 80. 
             {\it Right:} the same for the sub-sample of 3054 matches 
             with the number of good observations along scan at or above 
             the median. The thick blue line shows the best fit to 
             the Rayleigh distribution with parameter $\sigma= 1.38$. 
            }
    \label{f:distr_ngal}
\end{figure*}

   In order to check this hypothesis, we examined the parameter called
the number of good observations along scan direction (NgAL) provided in 
the \Gaia\ catalogue. NgAL varies from 2 to 1875 among the matches with 
the median value of 80. This parameter is proportional to the number of 
view crossings. We split the sample of matches into two equal sub-samples 
with NgAL below and above the median. The distributions among these 
sub-samples are indeed very different (Figure~\ref{f:distr_ngal}). The 
distribution in the sub-sample with  NgAL $\ge$ median fits reasonably well 
to the Rayleigh distribution with $\sigma=1.38$, but the sub-sample with 
NgAL $<$ median does not. This confirms our conjecture that estimation 
of parallaxes and proper motions is responsible for inflation of the 
reported uncertainties.

\begin{figure*}
    \includegraphics[width=0.40\textwidth]{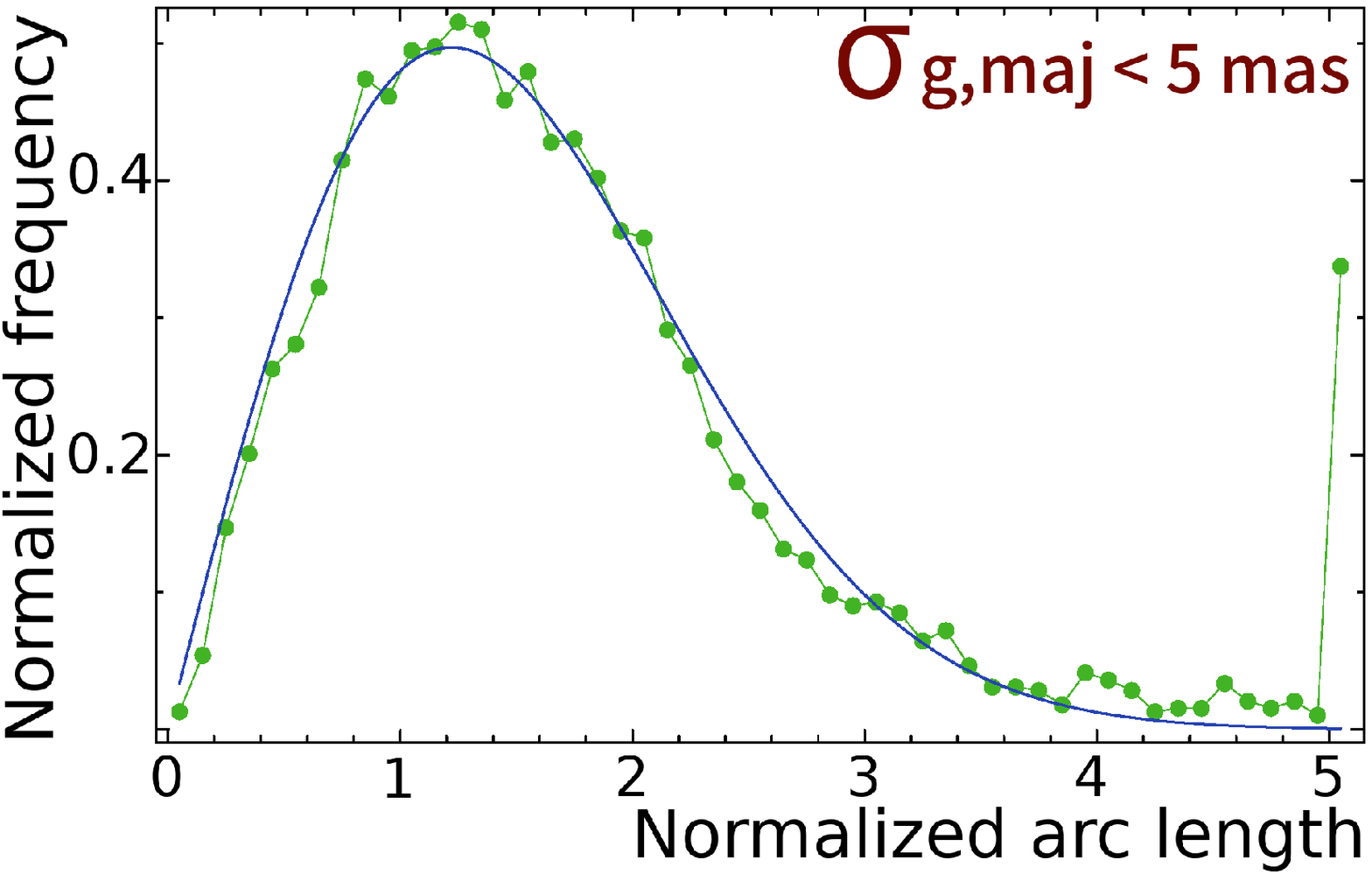}
    \includegraphics[width=0.40\textwidth]{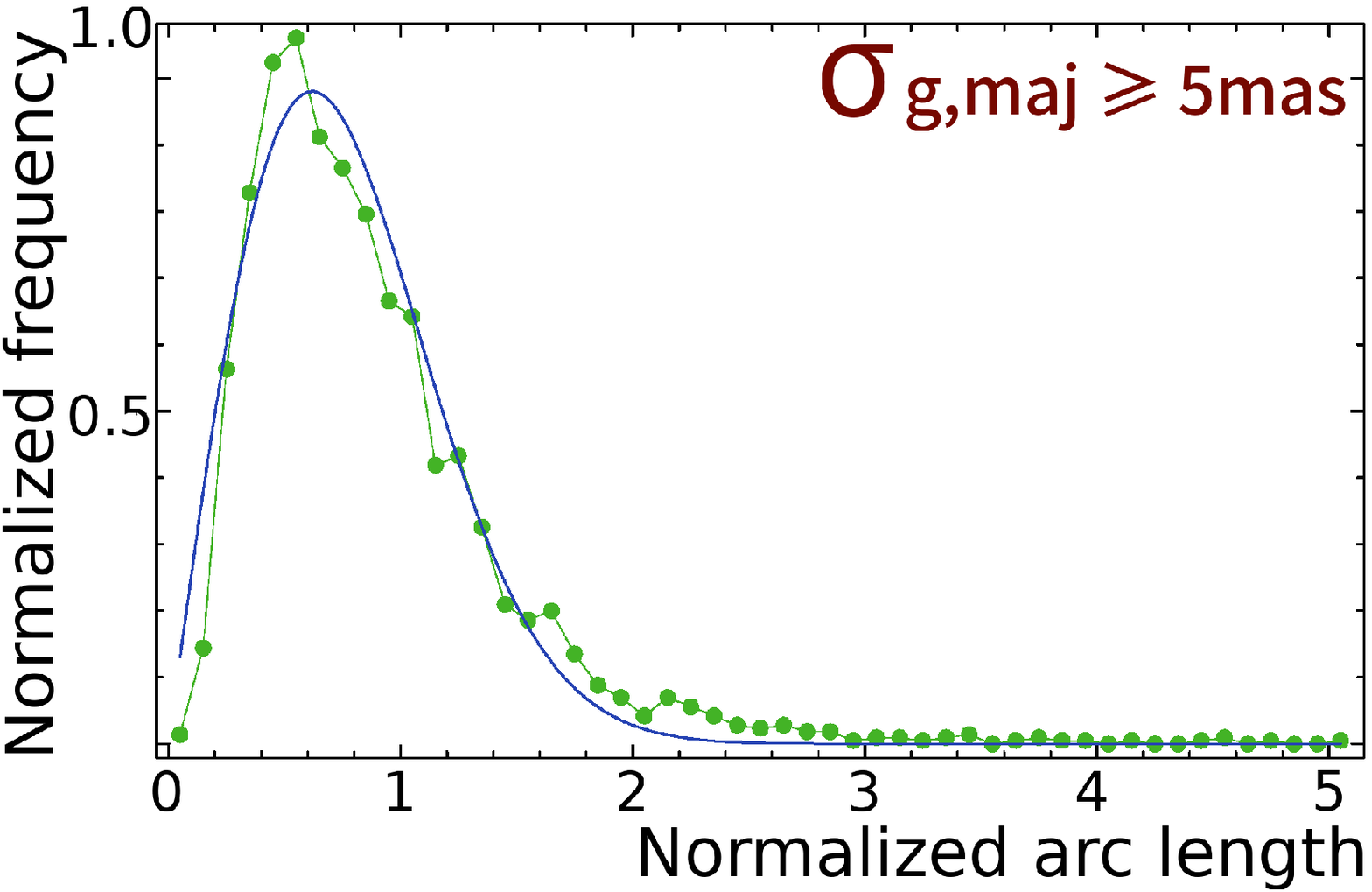}
    \caption{{\it Left:} the distribution of normalized arc lengths among 
             VLBI/\Gaia\ matches with $\sigma_{\rm g,maj} < 5$~mas after 
             the power transformation $\lambda = 0.829$. The blue line shows 
             the best fit of the Rayleigh distribution with $\sigma = 1.240$ to 
             the transformed distribution. {\it Right:} similar distribution 
             among VLBI/\Gaia\ matches with $\sigma_{\rm g,maj} \ge 5$~mas 
             after the power distribution $\lambda = 0.465$. The blue line 
             shows the best fit of the Rayleigh distribution with 
             $\sigma = 0.622$ to the transformed distribution.
            }
    \label{f:empi_distr}
\end{figure*}

   We sought for a simple smooth function close to the Rayleigh distribution
that can approximate the empirical distribution of normalized arc lengths.
Our further analysis showed that the distribution has different shape for 
{small and large Gaia position uncertainties.} We found that the distribution 
of normalized arc lengths $q$ of the sub-samples with Gaia semi-major 
error axes shorter and longer 5~mas can be represented as the Rayleigh 
distribution after applying the power law transformation $q^\lambda$ with 
different power and scale parameters. The Table~\ref{t:empi_distr} shows 
parameters of the transformation and Figure~\ref{f:empi_distr} illustrates 
the distributions of two sub-samples after the power law transformation and 
their best fit to the Rayleigh distributions. 

\begin{table}
   \caption{Parameters of the empirical model of normalized arc VLBI/\Gaia\ 
            for two ranges \Gaia\ semi-major error axes. The second 
            column shows the best fit to the power transformation parameter. 
            The third column shows the scaling parameter of the best fit to 
            the Rayleigh distribution after the power transformation. The last 
            column shows the root mean square (rms) of residuals after fitting.}
   \begin{center}
      \begin{tabular}{llll}
          \hline
          Range       & $\lambda$ & $\sigma$ & rms   \\
          \hline
          $<5$ mas    & 0.829     & 1.220    & 0.017 \\
          $\ge 5$ mas & 0.465     & 0.622    & 0.158 \\
          \hline
      \end{tabular}
   \end{center}
   \label{t:empi_distr}
\end{table}

We split the matches into the bulk subset which distribution obeys the 
the power-law Rayleigh functions and a subset of matches with the 
probability to belong to the bulk subset below some threshold, 
i.e. outliers. We consider the offsets from the bulk subset are due to the
random noise.

Since the probability $P(x>x_0) = e^{-\frac{x_0^2}{2\sigma^2}}$ for the 
Rayleigh distribution, we compute the probability for a given source 
to have the normalized power-law scaled arc length $q^\lambda$ equal 
or greater than a given value due to the random noise as 
$P(q^\lambda) = e^{-\frac{q^{2\lambda}}{2\sigma^2}}$, where 
$\sigma$ and $\lambda$ are parameters from Table~\ref{t:empi_distr}.

\section{Sources with statistically significant offsets}

\begin{figure}
    \centerline{\includegraphics[width=0.40\textwidth]{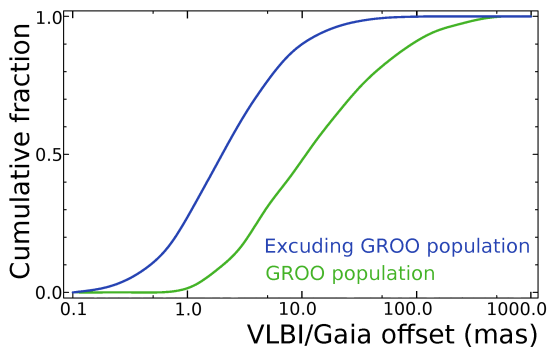}}
    \caption{\note{Cumulative distribution function of VLBI/\Gaia\ offsets 
             in logarithmic scale: green (lower) curve for the GROO 
             population and blue (upper) curve for remaining sources.}}
    \label{f:distr_offsets}
\end{figure}

  We consider an offset between VLBI and \Gaia\ positions statistically
significant if both the PFA is less than 0.0002 and the probability 
that the position offset is caused by the random noise (RNP) is less 
than 0.01. There are 384 matches (6\%) that satisfy these criteria. 
\Note{See their cumulative distribution in Figure~\ref{f:distr_offsets}}.
Table~\ref{t:roo} shows these sources. Table~3 with remaining 5671 matches 
with PFA $ < 0.0002$ and RNP $ \ge 0.01$ is given in the electronic 
attachment only. It should be noted the share of outliers among matches 
with the \Gaia\ DR1 solution is very close to the share of outliers 
with the \Gaia\ auxiliary quasar solution (also 6\%).

\begin{table*}
   \caption{The first 4 rows of the table of 384 VLBI/\Gaia\ matches with 
            statistically significant offsets: probability of false 
            association (PFA) less than 0.0002 and the random noise probability 
            (RNP) less than 0.01. The fifth column contains the normalized arc
            lengths, and two last columns contain positions of
            \Gaia\ minus VLBI over right ascensions, including $\cos\delta$ factor,
            and declination. The full table is available 
            in the electronic attachment.}
   \begin{tabular}{llllrrr}
       \hline
       \ntab{c}{VLBI ID} & \ntab{c}{Gaia ID} & \ntab{c}{PFA} & \ntab{c}{RNP} & \ntab{c}{q} & \ntab{c}{$d_\alpha$ (mas)} & \ntab{c}{$d_\delta$ (mas)} \\
       \hline
       RFC J0000$-$3221  & Gaia 2314315845817748992  & $ 4.47 \cdot 10^{-8} $ & $ 2.47 \cdot 10^{-22} $ &   20.78 &   -6.51 &  -0.83 \\
       RFC J0004$-$0802  & Gaia 2441584492826114432  & $ 3.58 \cdot 10^{-6} $ & $ 4.14 \cdot 10^{-03} $ &    4.73 &  -21.39 & -14.39 \\
       RFC J0005$+$3820  & Gaia 2880735411259458048  & $ 1.98 \cdot 10^{-7} $ & $ 5.03 \cdot 10^{-08} $ &   10.80 &    5.77 &  -3.43 \\
       RFC J0008$-$2339  & Gaia 2337107759788510464  & $ 2.01 \cdot 10^{-8} $ & $ 5.84 \cdot 10^{-06} $ &    8.84 &    1.17 &  -3.88 \\
       \ldots &&&&&& \\
       \hline
   \end{tabular}
   \label{t:roo}
\end{table*}

  A number of reasons may result in statistically significant offsets:
a)~errors in \Gaia\ positions; b)~errors in VLBI positions; c)~genuine 
radio optic offset (GROO). We will consider both \Gaia\ and VLBI errors that
led to significant offsets as failures of quality control rather than random 
errors. We investigated which objects are more common among the sources with 
statistically significant offsets and found three groups: 1)~sources with 
$\sigma_{\rm v,maj} > 5$~mas (a factor of 1.9 more common); 2)~sources 
brighter 17 mag (a factor of 2.7); and 3)~sources with 
$\sigma_{\rm g,maj} < 0.3$~mas (a factor of 1.6). The dominance of 
sources with position uncertainties greater 5~mas indicates a possible 
failure of the quality control of VLBI data analysis for {\it some} sources 
in that group. Position uncertainties greater than 5~mas are usually obtained 
when a source was close to the detection limit and too few observations were 
collected. The weaker the signal to noise ratio, the more chances that a wrong 
maximum in the delay resolution function will be selected. Errors in group 
delay that correspond to the wrong maximum are significantly greater than 
their formal uncertainty computed assuming a correct maximum was found.
The fewer observations, the more chances that a failure in fringe fitting 
will remain undetected. During past iterations of VLBI data analysis, a number 
of group delay estimates that correspond to an incorrect maximum in the delay 
resolution function were identified and fixed, which resulted in a change of 
source coordinate estimates. It is conceivable that not all such observations 
have been identified and eliminated. But such oversights in quality control 
affects noticeably only positions of sources with too few observations, usually
less than 20. The share of sources with 40 or less VLBI observations is 36\% 
among the objects with statistically significant offsets. That means that more 
than 2/3 matches with significant offsets cannot be affected by oversights 
in VLBI quality control.

  A greater share of optically bright sources with small \Gaia\ position 
errors favours a hypothesis that at least a part of objects with 
significant radio optic offsets are GROO: smaller position uncertainties 
make position offsets statistically more significant if they are real.

  Analysis of the group of sources with statistically significant 
offsets revealed there several gravitation lenses and a number of optically 
bright galaxies, but did not show any outstanding features that singles 
out these objects. The evidence collected so far supports the 
presence of the GROO population since observed significant radio/optic 
offsets cannot be explained only by failures in quality control. In order
to explain the phenomenon of GROO, additional information should be
examined. \citet{r:gaia_jets} investigated a connection between directions 
of AGN jets and offset directions. More studies focused on explanation of 
the GROO population are anticipated in the future.

\section{Summary}

  We explored offsets between \Gaia\ DR1 and VLBI positions. We used the 
secondary dataset for optical positions and recent VLBI solution 
rfc\_2016c based on analysis of all available observations 
suitable for absolute astrometry collected since 1980 through July 2016.
We have found 6055 matched AGNs using the criterion set on their 
arc lengths, such that the mathematical expectation of the number of 
spurious matches in this sample is less than one object. When we 
used the \Gaia\ auxiliary quasar solution, we were able to reproduce 
closely results of \citet{r:gaia_icrf2}. 

  Comparison of \Gaia\ DR1 and VLBI solutions revealed the following.
\begin{itemize}
   \item The median position offset is 2.2~mas --- very close to the
         median semi-major axis of the error ellipse of \Gaia\ positions
         in the entire dataset.

   \item The median semi-major axis of the error ellipse of \Gaia\ positions
         among the matches, 2.1~mas, is a factor of 4 greater than the 
         median semi-major axis of the error ellipse of VLBI positions.

   \item The distribution of normalized arc lengths is significantly 
         non-Rayleighian. We found evidence that the analysis
         strategy implemented in \Gaia\ DR1 disfavored sources with
         negligible parallaxes and proper motions, which inflated their
         uncertainties.
         
   \item There exits a population of sources with offsets statistically
         significant at the 99\% confidence level (6\% of the matches).
         We admit that some these objects may have statistically 
         significant offset due to failures in quality control in both 
         VLBI and \Gaia\, but certainly, not all: at maximum 1/3. 
         An increased share of optically bright objects with small position 
         uncertainties in this population suggests that some these 
         objects have genuine radio optical offsets (GROO).

\end{itemize}

   The emission center in optic and in radio may not always coincide
for a number of reasons. Firstly, the centroid of the core may be shifted 
with frequency \citep[e.g.,][]{r:L98,r:Kovalev_cs_2008}. Secondly, unaccounted 
radio structure may cause an offset of the reference point with respect to 
the jet base, although such a shift is usually below 1~mas. Thirdly, as 
\citet{r:v2m} shown, there exist interacting galaxies within 
an optically weaker component hosting a bright radio source. In the 
era of ground optical astrometry, a study of such objects was limited 
to pairs at least $1''$ apart. \Gaia\ astrometry has a potential to find 
such objects separated at milliarcsecond level. Finally, the presence of 
bright components along the jet may shift the optic centroid. At the 
moment, little is known about properties of jets at milliarcsecond 
scales in optic wavelengths. Investigation of the GROO population opens 
a new window into study of AGNs. 

  We should stress that this analysis is based on \Gaia\ DR1 secondary 
dataset and we expect statistics of comparison VLBI positions and future 
\Gaia\ releases will be significantly different because of anticipated 
changes in data analysis strategy of \Gaia\ observations.

\section*{Acknowledgments}

It is our pleasure to thank Alexey Butkevich, Sergei Klioner, Alexandr 
Plavin, and Eduardo Ros for fruitful discussions. \Note{We are very grateful to
Lennart Lindegren for a detailed referee report and suggestions that helped 
us greatly to improve the manuscript and fix an error in numerical 
tables.} This work is supported by the Russian Science Foundation grant 
\mbox{16--12--10481}.

\bibliographystyle{mnras}
\bibliography{rfc_gaia}

% Don't change these lines
\bsp    % typesetting comment
\label{lastpage}
\end{document}